# Geodesic learning

Arnab Barua [a], Haralampos Hatzikirou [b,c], Sumiyoshi Abe [d,e,f *]

[a] *Tata Institute of Fundamental Research, Hyderabad 500046, India*
[b] *Technische Univesität Dresden, Center for Information Services and High Performance Computing, Nöthnitzer Straße 46, Dresden 01187, Germany*
[c] *Mathematics Department, Khalifa University, P.O. Box: 127788, Abu Dhabi, UAE*
[d] *Department of Physics, College of Information Science and Engineering, Huaqiao University, Xiamen 361021, China*
[e] *Institute of Physics, Kazan Federal University, Kazan 420008, Russia*
[f] *Department of Natural and Mathematical Sciences, Turin Polytechnic University in Tashkent, Tashkent 100095, Uzbekistan*

* Corresponding author: suabe@sf6.so-net.ne.jp

ABSTRACT

Learning is a fundamental characteristic of living systems, enabling them to comprehend their environments and make informed decisions. These decision-making processes are inherently influenced by available information about their surroundings and specific objectives. There is an intriguing perspective is that each process is highly efficient under a given set of conditions. A key question, then, is how close to optimality it is or how efficient it is under given conditions. Here, the concept of *geodesic learning* as the optimal reference process, with which each process can be compared, is introduced and formulated on the basis of geometry. The probability distribution describing the state of the composite system consisting of the environment, termed the *information bath*, and a decision-maker is described by use of the entropic quantities. This enables one to study the system in analogy with thermodynamics. Learning processes are expressed as the changes of parameters contained in the distribution. For a geometric interpretation of the processes, the manifold endowed with the Fisher-Rao metric as the Riemannian metric is considered. This framework allows one to conceptualize the optimality of each process as a state change along a geodesic curve on the manifold, which gives rise to geodesic learning. Then, the bivariate Gaussian model is presented, and the processes of geodesic learning and adaptation are analyzed for illustrating this approach.

## 1. Introduction

Living agents or organisms survive by inferring their surroundings through learning processes. Suppose a living object to be eventually put in a new environment. Then, it will start changing its initial state through multiple stages of cognitive/sensing mechanisms to gain information on the environment for survival. This can be viewed as an adaptation process [1]. In reality, different organisms use different processes to decode information contained in the environment. This may depend on the phylogenetic tree in evolution, in general [2,3]. In the current situation, our knowledge of such processes is still limited, microscopically. Therefore, to advance understanding them, it is natural to describe, in some macroscopic way, each survival process of a living object in terms of a learning process of the object as a decision-maker. It may be of particular interest if such a description is made without recourse to the perplexing details of microscopic processes. Although these processes in the real world are often considered to be highly efficient, it is generally unknown if they are "optimal" (or optimal in what sense). This is in marked contrast to thermodynamics, where the Carnot cycle plays a distinguished role. To improve this situation, it is of importance to characterize the optimal process as a reference process, with which each real process can be compared.

In this paper, we present one such description. For it, we base our discussion on the thermodynamics analogy of learning and geometric deliberation. We consider the entropy associated with learning and introduce the concepts of the "information bath" and the "specific information", which are analogs of the heat bath and specific heat in thermodynamics, respectively. There, the specific information is defined by the entropy fluctuation. The state of the combined system of the information bath and a decision-maker is described by a joint probability distribution characterized by a set of parameters. Therefore, once the functional form of the joint distribution is specified, a learning process can be seen as the change of the parameters. As a guiding principle, we employ the view that the process taken by the decision-maker is optimal. To geometrically formulate the optimality, we consider the geodesics on the Riemannian manifold endowed with the Fisher-Rao metric [4] as the Riemannian metric. Since the Fisher-Rao metric is written in terms of the parameters as the local coordinate on the manifold, the geodesic curves are described by the macroscopic quantities such as the variances and correlations obtained by integrating out (i.e. eliminating) the microscopic variables of the information bath and the decision-maker. We define the learning process along the geodesic curves on the manifold, which is referred to here as *geodesic learning*. We do not assume any kinetic equation for the distribution. The system is not constrained by *a priori* dynamics and evolves following its own strategy for geometric optimality. We discuss the entropic measures of the system to make the thermodynamic analogy manifest and describe learning as the decrease in the subentropy of the decision-maker. To illustrate the idea, we present the bivariate Gaussian model and analyze it, in detail. In it, each of the information bath and the decision-maker is simply described by a single variable. In other words, only the minimal environment (i.e. the microenvironment [5]) primarily relevant to the learning considered is employed as the information bath. Accordingly, the combined system here may not be an isolated one but yet an open system in nonequilibrium states. We evaluate the joint entropy, subentropies and correlation between the information bath and the decision-maker. We show how learning can be seen as the analog of cooling. Analyzing geodesic evolution of the entropies and correlation, we show how the model exhibits the processes of geodesic learning and associated adaptation, simultaneously. Although our purpose here is not to discuss specific real-world learning processes but to formulate, based on the geometric concepts and thermodynamic analogies, a new general framework for describing the optimal process, the results are in fact found to be natural.

The rest of this paper is organized as follows. In Section 2, the idea is described about learning in the thermodynamic analogy. The concepts of the information bath and the specific information are introduced. Then, a discussion is developed about geodesic learning defined on the Riemannian manifold with the Fisher-Rao metric. In Section 3, the bivariate Gaussian model is presented as an example and its detailed analysis is performed. A general condition for learning is derived. Then, the explicit form of the geodesic equation is presented. The entropies of the composite system and the subsystems are evaluated for geodesic learning. The associated adaptation process of the decision-maker to the information bath is discussed. In Section 4, the work is summarized and a comment is made on possible applications of the present approach to problems in other fields such as psychology and epigenetics.

## 2. Information bath and geodesic learning

In equilibrium thermodynamics, what fundamentally relevant are the dynamical variables (and corresponding coarse-grained ones), and the distribution does not play any explicit dynamical role. In nonequilibrium, however, this situation radically becomes different. Both the variables and the distribution are dynamical. (This doubled structure may remind one of that in quantum thermodynamics: the observables and the Hilbert space. Accordingly, there appears a possibility of a variety of baths, not limited to the heat bath. The so-called energy bath associated with the "isoenergetic process" is one such example [6].) Learning processes take place typically under nonequilibrium circumstances. This situation is in analogy with thermodynamics of a subsystem in contact with the heat bath. However, what is exchanged is not heat but information. This motivates us to introduce the concept of the "information bath".

Let us consider a bipartite system consisting of a decision-maker and the information bath surrounding it, and denote their state variables by $\mathbf{X} = (X_1, X_2, ..., X_l)$ and $\mathbf{Y} = (Y_1, Y_2, ..., Y_m)$, respectively. These are random variables, which effectively describe sensing of the information flow. The state of the composite system is supposed to be given by a certain normalized probability distribution $p(\mathbf{x}, \mathbf{y}; \Theta)$, where $\mathbf{x} = (x_1, x_2, ..., x_l)$ and $\mathbf{y} = (y_1, y_2, ..., y_m)$ are the realizations of $\mathbf{X}$ and $\mathbf{Y}$, and $\Theta = (\theta^1, \theta^2, ..., \theta^n)$ stands for the set of parameters contained in the distribution

The state of the decision-maker is described by the marginal distribution

$$p(\mathbf{x}; \Theta) = \int d^m \mathbf{y} \; p(\mathbf{x}, \mathbf{y}; \Theta) \,. \tag{1}$$

The decision-maker is not independent of the environment, in general. That is, the conditional distribution $p(\mathbf{y};\Theta\,|\,\mathbf{x}) = p(\mathbf{x},\mathbf{y};\Theta)\,/\,p(\mathbf{x};\Theta)$ may depend on $\mathbf{x}$. In Eq. (1), the notation in the case of the continuous random variables is employed, but the expression corresponding to the discrete variables is obvious.

Let us consider the subentropy of the decision-maker. In the case of the continuous random variables, it is the differential entropy [7]

$$S[\mathbf{X}] = -\int d^l\mathbf{x}\; p(\mathbf{x};\Theta) \ln p(\mathbf{x};\Theta), \tag{2}$$

provided that, instead of bits, the unit of the natural base is used here for the later convenience. In contrast to the Shannon entropy in the case of discrete random variables, the differential entropy can take negative values. In addition, it is desirable to make the random variables dimensionless, and it is in fact assumed to be so in the subsequent discussion. Then, the following celebrated composition rule holds:

$$S[\mathbf{X},\mathbf{Y}] = S[\mathbf{Y}\,|\,\mathbf{X}] + S[\mathbf{X}] = S[\mathbf{X}\,|\,\mathbf{Y}] + S[\mathbf{Y}], \tag{3}$$

where $S[\mathbf{X},\mathbf{Y}]$ is the entropy of the joint distribution $p(\mathbf{x},\mathbf{y};\Theta)$, and $S[\mathbf{Y}\,|\,\mathbf{X}]$ is the conditional entropy given by

$$S[\mathbf{Y}\,|\,\mathbf{X}] = \int d^l\mathbf{x}\; p(\mathbf{x};\Theta)\; S[\mathbf{Y}\,|\,\mathbf{x}) \tag{4}$$

with $S[\mathbf{Y}\,|\,\mathbf{x})$ being

$$S[\mathbf{Y}\,|\,\mathbf{x}) = -\int d^m\mathbf{y}\; p(\mathbf{y};\Theta\,|\,\mathbf{x}) \ln p(\mathbf{y};\Theta\,|\,\mathbf{x})\,. \tag{5}$$

$S[\mathbf{X}\,|\,\mathbf{Y}]$ is defined in a similar way. All of the quantities in Eqs. (2)-(5) are the functions of $\Theta$.

It is also important to clarify the correlation between the decision-maker and the environment as the information bath. An information-theoretic measure of correlation may be the mutual information [7]

$$I[\mathbf{X};\mathbf{Y}] = S[\mathbf{X}] + S[\mathbf{Y}] - S[\mathbf{X},\mathbf{Y}]\,, \tag{6}$$

which is also rewritten as follows: $I[\mathbf{X};\mathbf{Y}] = S[\mathbf{X}] - S[\mathbf{X}\,|\,\mathbf{Y}] = S[\mathbf{Y}] - S[\mathbf{Y}\,|\,\mathbf{X}]$. This quantity is nonnegative and vanishes if and only if $\mathbf{X}$ and $\mathbf{Y}$ are independent (i.e. uncorrelated).

To establish an analogy to thermodynamics, we consider the entropy fluctuation although, strictly speaking, such a concept is not usually discussed in thermodynamics. We define the fluctuation of the decision-maker subentropy as the variance of the information content:

$$\left(\Delta S[\mathbf{X}]\right)^2 = \int d^l\mathbf{x}\, p(x;\Theta)\left[-\ln p(x;\Theta)\right]^2 - \left\{\int d^l\mathbf{x}\, p(x;\Theta)\left[-\ln p(x;\Theta)\right]\right\}^2, \tag{7}$$

where the second term on the right-hand side is equal to $-S^2[\mathbf{X}]$. To clarify the meaning of this quantity, it is convenient to evaluate it for the distribution in the canonical ensemble in statistical mechanics, that is, $p_i = Z^{-1}(\beta)\exp(-\beta\,\varepsilon_i)$, where $\varepsilon_i$ is the $i$-th value of the energy of the subsystem, $\beta$ the inverse temperature (with Boltzmann's constant, $k_B$, being set equal to unity, here), and $Z(\beta)$ the canonical partition function given by $Z(\beta) = \sum_i \exp(-\beta\,\varepsilon_i)$. In this case, the entropy fluctuation is found to be $\left(\Delta S\right)^2 = \sum_i p_i (-\ln p_i)^2 - \left(\sum_i p_i [-\ln p_i]\right)^2 = \beta^2 (\Delta E)^2$, where $(\Delta E)^2$ is the variance of the energy. In the canonical ensemble theory, the energy fluctuation is directly related to the specific heat, $C$, as $(\Delta E)^2 = C / \beta^2$ ($k_B \equiv 1$). This fact naturally leads to the notion of the *specific information* defined by the entropy fluctuation in Eq. (7), which is an analog of the specific heat and quantifies the decision-maker's capability of receiving information from the information bath, which may measure robustness of the decision-maker.

Now, let us formulate learning process. Our idea is to geometrically describe it. Once the functional form of the distribution is specified, learning is the state change that can be represented as the change of the parameters (on this point, see the comments in the next paragraph). That is,

$$p(\mathbf{x}, \mathbf{y}; \Theta) \to p(\mathbf{x}, \mathbf{y}; \Theta + d\Theta) . \tag{8}$$

To quantify this change, we use the squared distance between the two

$$ds^2 = g_{ij}(\Theta)\, d\theta^i\, d\theta^j , \tag{9}$$

where $g_{ij}(\Theta)$ stands for the Fisher-Rao metric [4] given in the form as follows:

$$g_{ij}(\Theta) = \iint d^l\mathbf{x}\, d^m\mathbf{y}\; p(\mathbf{x},\mathbf{y};\Theta) \frac{\partial \ln p(\mathbf{x},\mathbf{y};\Theta)}{\partial \theta^i} \frac{\partial \ln p(\mathbf{x},\mathbf{y};\Theta)}{\partial \theta^j}. \tag{10}$$

In Eq. (9), the summation convention is understood for the repeated upper and lower indices: e.g. $a_{ij} b^i \equiv \sum_{i=1}^{n} a_{ij} b^i$. Equation (10) has its entropic origin: it is the relative entropy (the Kullback-Leibler divergence) [7], quantifying the difference between $p(\mathbf{x},\mathbf{y};\Theta)$ and $p(\mathbf{x},\mathbf{y};\Theta + d\Theta)$ that are infinitesimally separated from each other. Thus, a process can be expressed as "motion" on the *n*-dimensional Riemannian manifold endowed with the Fisher-Rao metric as the Riemannian metric. $\Theta = (\theta^1, \theta^2, ..., \theta^n)$ plays a role of a local coordinate. *s* in Eq. (9) is the arc length and is termed the affine parameter. It corresponds to the proper time in relativity, the lapse of which is measured by the clock attached to a moving object. In the present case, the clock ticks according to the state change. Therefore, *s* may be thought of as the *internal time* and is different from the conventional coordinate time.

In the above, we have considered the situation that the functional form of the distribution is specified. We note that this is not a stringent restriction on the discussion for the change of the distribution during a learning process. Introducing parameters as the coordinate variables, one may cover a wide class of such processes. For example, changes between the power-law and exponential distributions or between unimodal and multimodal distributions can be described through introductions of relevant varying parameters. Therefore, such changes are understood to be included in Eq. (8). Once those parameters reach their values realizing a specific distribution of interest, then the subsequent changes are taken place only on the submanifold associated with those fixed values of the parameters.

A process is "motion" along a curve $\Theta(s) = (\theta^1(s), \theta^2(s), ..., \theta^n(s))$ on the manifold. We define the learning condition for the decision-maker as follows:

$$\frac{d\,S[\mathbf{X}]}{d\,s} < 0\,. \tag{11}$$

In general, this condition can be viewed as antidiffusion.

Of particular interest is the case when a curve is geodesic. Then, the parameters satisfy the following equation [8]:

$$\frac{d^2\theta^i}{d\,s^2} + \Gamma^i_{jk}\frac{d\,\theta^j}{d\,s}\frac{d\,\theta^k}{d\,s} = 0\,, \tag{12}$$

where $\Gamma^i_{jk}$ is the Christoffel symbol given by

$$\Gamma^i_{jk} = \frac{1}{2}g^{ih}\left(\frac{\partial g_{hk}}{\partial\theta^j} + \frac{\partial g_{jh}}{\partial\theta^k} - \frac{\partial g_{jk}}{\partial\theta^h}\right). \tag{13}$$

Given initial/final conditions, the process along the geodesic curve is interpreted to be geometrically optimal. We call such a process under the condition in Eq. (11) *geodesic learning*.

Closing this section, we wish to emphasize the following point. In the present approach, learning process is described in terms only of $\Theta$. In general, $\Theta$ can be expressed as the expectation values of quantities that are the functions of $(\mathbf{X}, \mathbf{Y})$. In other words, $(\mathbf{X}, \mathbf{Y})$ and $\Theta$ are the microvariables and macrovariables, respectively. Therefore, our approach utilizes the effective variables based on such scale separation. The internal time $s$ emerges as the state change of the system itself. It is noted that no kinetic equations are assumed for the distributions. As emphasized in Section 1, the system evolves following the strategy for geometric optimality without being constrained by *a priori* dynamics.

## 3. Bivariate Gaussian model

To illustrate geodesic learning, we present the bivariate Gaussian model. Here, the random variables associated with the decision-maker and the information bath are $X$ and $Y$, respectively. Thus, the information bath is as small as the decision-maker and is interpreted as the minimal part of the environment (i.e. the microenvironment) relevant to learning of the decision-maker. Thus, the composite system $(X, Y)$ is not isolated but open. As will be seen, a crucial point distinguishing between the decision-maker $X$ and the information bath $Y$ is the *time-scale separation* between them: they are dynamically the fast and slow degrees of freedom, respectively. The state of the composite system is assumed to be described by the bivariate Gaussian distribution

$$p(x, y; \Theta) = N \exp\left(-\theta^1 y^2 + 2\theta^2 xy - \theta^3 x^2\right), \tag{14}$$

where the support of the distribution is the whole $xy$-plane, $N$ is the normalization factor

$$N = \sqrt{\frac{\theta^1\theta^3 - (\theta^2)^2}{\pi^2}}\,, \tag{15}$$

and the parameters $\Theta = (\theta^1, \theta^2, \theta^3)$ are required to satisfy

$$\theta^1, \theta^3 > 0\,, \tag{16}$$

$$\theta^1\theta^3 - (\theta^2)^2 > 0\,. \tag{17}$$

The terms linear in $x$ and $y$ could also be included in the exponential in Eq. (14), but the present form is simple and sufficient for our purpose, here.

The marginal and conditional distributions are

$$p(x;\Theta) = N\sqrt{\frac{\pi}{\theta^1}}\exp\left(-\frac{\pi^2 N^2}{\theta^1}x^2\right), \tag{18}$$

$$p(y;\Theta) = N\sqrt{\frac{\pi}{\theta^3}}\exp\left(-\frac{\pi^2 N^2}{\theta^3}y^2\right), \tag{19}$$

$$p(x;\Theta\,|\,y) = \sqrt{\frac{\theta^3}{\pi}}\exp\left[-\theta^3\left(x - \frac{\theta^2}{\theta^3}y\right)^2\right], \tag{20}$$

$$p(y;\Theta\,|\,x) = \sqrt{\frac{\theta^1}{\pi}}\exp\left[-\theta^1\left(y - \frac{\theta^2}{\theta^1}x\right)^2\right]. \tag{21}$$

These obey the Bayes rule: $p(x, y;\Theta) = p(y;\Theta\,|\,x)\,p(x;\Theta) = p(x;\Theta\,|\,y)\,p(y;\Theta)$. Although the present work is not concerned with Bayesian learning, yet we mention that

if $p(x;\Theta)$ and $p(x;\Theta\,|\,y)$ are regarded as the prior and the posterior in the Bayesian, then their variances (i.e. the squared widths) are seen to shrink from $1/[2\,\theta^3 - 2(\theta^2)^2/\theta^1]$ to $1/(2\theta^3)$, respectively.

As emphasized in the preceding section, the parameters are expressed in terms of the expectation values as follows: $\left\langle X^2 \right\rangle = \theta^1/(2\pi^2 N^2)$, $\left\langle Y^2 \right\rangle = \theta^3/(2\pi^2 N^2)$ and $\left\langle XY \right\rangle = \theta^2/(2\pi^2 N^2)$, where $\left\langle Q(X,Y) \right\rangle \equiv \iint dx\,dy\,Q(x,y)\,p(x,y;\Theta)$ and $N$ is given in Eq. (15). Since $\left\langle X \right\rangle = 0$ and $\left\langle Y \right\rangle = 0$, these three quantities are directly equal to the variances and covariance, respectively: $\left(\Delta X\right)^2 = \left\langle X^2 \right\rangle - \left\langle X \right\rangle^2 = \left\langle X^2 \right\rangle$, $\left(\Delta Y\right)^2 = \left\langle Y^2 \right\rangle - \left\langle Y \right\rangle^2 = \left\langle Y^2 \right\rangle$ and $C(X,Y) = \left\langle X\,Y \right\rangle - \left\langle X \right\rangle \left\langle Y \right\rangle = \left\langle X\,Y \right\rangle$. Reversely, we have $\theta^1 = 2\pi^2 N^2 (\Delta X)^2$, $\theta^3 = 2\pi^2 N^2 (\Delta Y)^2$ and $\theta^2 = 2\pi^2 N^2 C(X,Y)$ together with $N^2 = 1/(4\pi^2 \det\Gamma)$, where $\Gamma$ is the $2\times 2$ covariance matrix with the elements $\Gamma_{11} = (\Delta X)^2$, $\Gamma_{22} = (\Delta Y)^2$ and $\Gamma_{12} = \Gamma_{21} = C(X,Y)$. The joint entropy, subentropies and conditional entropies are found to be

$$S[X,Y] = 1 - \ln N = 1 + \ln(2\pi) + \frac{1}{2}\ln(\det\Gamma)\,, \tag{22}$$

$$S[X] = \frac{1}{2} + \frac{1}{2}\ln\frac{\theta^1}{\pi N^2} = \frac{1}{2}[1 + \ln(2\pi)] + \frac{1}{2}\ln(\Delta X)^2\,, \tag{23}$$

$$S[Y] = \frac{1}{2} + \frac{1}{2}\ln\frac{\theta^3}{\pi N^2} = \frac{1}{2}[1 + \ln(2\pi)] + \frac{1}{2}\ln(\Delta Y)^2\,, \tag{24}$$

$$S[X\,|\,Y] = \frac{1}{2}(1+\ln\pi) - \frac{1}{2}\ln\theta^3 = \frac{1}{2}[1-\ln(2\pi)] - \frac{1}{2}\ln(\Delta Y)^2 - \ln N\,, \tag{25}$$

$$S[Y\,|\,X] = \frac{1}{2}(1+\ln\pi) - \frac{1}{2}\ln\theta^1 = \frac{1}{2}[1-\ln(2\pi)] - \frac{1}{2}\ln(\Delta X)^2 - \ln N\,, \tag{26}$$

which satisfy the composition law, $S[X,Y] = S[Y\,|\,X] + S[X] = S[X\,|\,Y] + S[Y]$, as they should do.

To quantify the correlation between $X$ and $Y$, we evaluate the mutual information [see Eq. (6)]. From Eqs. (22)-(24), it is calculated to be $I[X;Y] = (1/2)\ln\left[\theta^1\theta^3/(\pi^2 N^2)\right] = (1/2)\ln\left[(\Delta X)^2(\Delta Y)^2/\det\Gamma\right]$. This quantity, however, does not distinguish between correlation and anticorrelation (i.e., the signs of $\theta^2$), unlike the covariance.

The specific information of the decision-maker is the entropy fluctuation and is found to be

$$\left(\Delta S[X]\right)^2 = \int dx\, p(x;\Theta)\left[-\ln p(x;\Theta)\right]^2 - S^2[X] = \frac{1}{2}\,, \tag{27}$$

implying that the capability of information gain remains constant. This is due to a salient feature of the Gaussianity.

It is of interest to consider the analog of the thermodynamic internal energy. Taking into account the quadratic nature in the exponential in Eq. (18), we assume $X^2$ to be an analog of the energy variable (up to an unknown positive multiplicative constant, but such an ambiguity does not change the subsequent result). Then, the analog of the internal energy is $U = \left\langle X^2 \right\rangle = \theta^1 / (2\pi^2 N^2)$, by which Eq. (23) is rewritten as $S[X] = [1 + \ln(2\pi U)]/2$. This allows us to use the thermodynamic relation to define the analog of temperature $T$ as $1/T = \partial S[X]/\partial U$. From it, we have $U = T/2$, analogous to the law of equipartition of energy. This leads to the fact that the specific information corresponding to the specific heat is $\partial U / \partial T = 1/2$, consistently with Eq. (27). Thus, we obtain

$$T = \frac{\theta^1}{\pi^2 N^2} \tag{28}$$

as the analog of the temperature.

Let us address ourselves to describing learning of the decision-maker. The learning condition $d S[X]/d s < 0$ [see Eq. (11)] in the present case amounts to

$$(\theta^2)^2 \dot{\theta}^1 + (\theta^1)^2 \dot{\theta}^3 > 2\theta^1 \theta^2 \dot{\theta}^2, \tag{29}$$

where the over-dot denotes the differentiation with respect to *s*. In addition, since $S[X] = (1 + \ln \pi + \ln T)/2$, the learning condition gives rise to

$$\dot{T} < 0. \tag{30}$$

This is a natural result from the viewpoint that learning is the analog of cooling.

So far, we have mainly considered the thermodynamic analogies. Next, let us discuss geodesic learning. The Riemannian manifold associated with the present bivariate Gaussian model is 3-dimensional. The elements of the Fisher-Rao metric tensor in the local coordinate $(\theta^1, \theta^2, \theta^3)$ are found to be

$$g_{11} = \frac{(\theta^3)^2}{2\pi^4 N^4}, \qquad g_{12} = -\frac{\theta^2\theta^3}{\pi^4 N^4}, \qquad g_{13} = \frac{(\theta^2)^2}{2\pi^4 N^4},$$
$$g_{22} = \frac{\theta^1\theta^3 + (\theta^2)^2}{\pi^4 N^4}, \qquad g_{23} = -\frac{\theta^1\theta^2}{\pi^4 N^4}, \qquad g_{33} = \frac{(\theta^1)^2}{2\pi^4 N^4}, \tag{31}$$

together with the symmetry $g_{ij} = g_{ji}$. Furthermore, the elements of the Christoffel symbol are calculated to be

$$\Gamma^1_{11} = -\frac{\theta^3}{\pi^2 N^2}, \qquad \Gamma^1_{12} = \frac{\theta^2}{\pi^2 N^2}, \qquad \Gamma^1_{22} = -\frac{\theta^1}{\pi^2 N^2},$$
$$\Gamma^1_{13} = \Gamma^1_{23} = \Gamma^1_{33} = 0, \qquad \Gamma^2_{11} = \Gamma^2_{33} = 0, \qquad \Gamma^2_{12} = -\frac{\theta^3}{2\pi^2 N^2},$$
$$\Gamma^2_{13} = \frac{\theta^2}{2\pi^2 N^2}, \qquad \Gamma^2_{22} = \frac{\theta^2}{\pi^2 N^2}, \qquad \Gamma^2_{23} = -\frac{\theta^1}{2\pi^2 N^2},$$
$$\Gamma^3_{11} = \Gamma^3_{12} = \Gamma^3_{13} = 0, \qquad \Gamma^3_{22} = -\frac{\theta^3}{\pi^2 N^2}, \qquad \Gamma^3_{23} = \frac{\theta^2}{\pi^2 N^2},$$
$$\Gamma^3_{33} = -\frac{\theta^1}{\pi^2 N^2}, \tag{32}$$

also together with the symmetry $\Gamma^i_{jk} = \Gamma^i_{kj}$. Therefore, the geodesic equation in Eq. (12) are explicitly written down as follows:

$$\ddot{\theta}^1 + \frac{1}{\pi^2 N^2}[-\theta^3(\dot{\theta}^1)^2 + 2\theta^2\dot{\theta}^1\dot{\theta}^2 - \theta^1(\dot{\theta}^2)^2] = 0, \tag{33}$$

$$\ddot{\theta}^2 + \frac{1}{\pi^2 N^2}[\theta^2(\dot{\theta}^2)^2 - \theta^3\dot{\theta}^1\dot{\theta}^2 + \theta^2\dot{\theta}^1\dot{\theta}^3 - \theta^1\dot{\theta}^2\dot{\theta}^3] = 0, \tag{34}$$

$$\ddot{\theta}^3 + \frac{1}{\pi^2 N^2}[-\theta^3(\dot{\theta}^2)^2 + 2\theta^2\dot{\theta}^2\dot{\theta}^3 - \theta^1(\dot{\theta}^3)^2] = 0\,. \tag{35}$$

These are coupled nonlinear equations, and it is a numerical problem to analyze them, in general.

The learning condition is concerned with the first-order derivative of $S[X] = S[X](s)$ with respect to *s*, whereas the geodesic equation is of the second-order. This suggests that the second-order derivative of $S[X]$ along the geodesic should give insight into geodesic learning. In fact, we obtain the following result directly coming from the geodesic equation:

$$\ddot{S}^*[X] = \frac{[W(\theta^1,\theta^2)]^2}{2(\theta^1)^2\pi^2 N^2}\,, \tag{36}$$

where the asterisk indicates that the quantity is evaluated along the geodesic, and $W(\theta^1,\theta^2)$ stands for the Wronskian, $W(\theta^1,\theta^2) = \theta^1\dot{\theta}^2 - \theta^2\dot{\theta}^1$. Therefore, we find that the second-order derivative of the subentropy of the decision-maker is nonnegative along geodesics. Later, we will see that actually

$$\ddot{S}^*[X] > 0 \tag{37}$$

should hold for the geodesic learning.

Below, we present both the result of numerical analysis of the present model and an exact solution in a special case.

**(3-1) Numerical analysis:** Upon analyzing the entropies, variance and covariance based on Eqs. (33)-(35), it is natural to impose asymmetry between *X* and *Y*: $S^*[X](0)$ is fairly larger than $S^*[Y](0)$. This reflects the situation of interest that the decision-maker starts learning, poorly knowing about the information bath (i.e. the environment), initially. From Eqs. (23) and (24), we see that

$$\theta^1(0) >> \theta^3(0) \tag{38}$$

should hold.

In Figs. 1 and 2, we present the plots of the numerical results of $\theta^i$'s, $S^*[X]$, $S^*[Y]$, $C^*(X,Y)$ and $\ddot{S}^*[X]$ with respect to the affine parameter $s$ under two different initial conditions. The learning condition in Eq. (29) as well as Eqs. (16) and (17) are ascertained. The processes are slow since the logarithmic internal time, $\log s$, is employed in the figures.

In Fig. 1, the initial correlation is positive. The result shows how adaptation of the state of the decision-maker to that of the information bath along the antisigmoidal curve: $S^*[X]$ decreases much more rapidly than $S^*[Y]$, showing the time-scale separation between the decision-maker and the information bath. $\ddot{S}^*[X,Y]$ has a large peak in the region where $S^*[X]$ rapidly decreases, and then rapidly decrease as the decision-maker comes to get adapted to the information bath as the environment. $C^*(X,Y)$ exhibits the sigmoidal behavior, showing another aspect of adaptation.

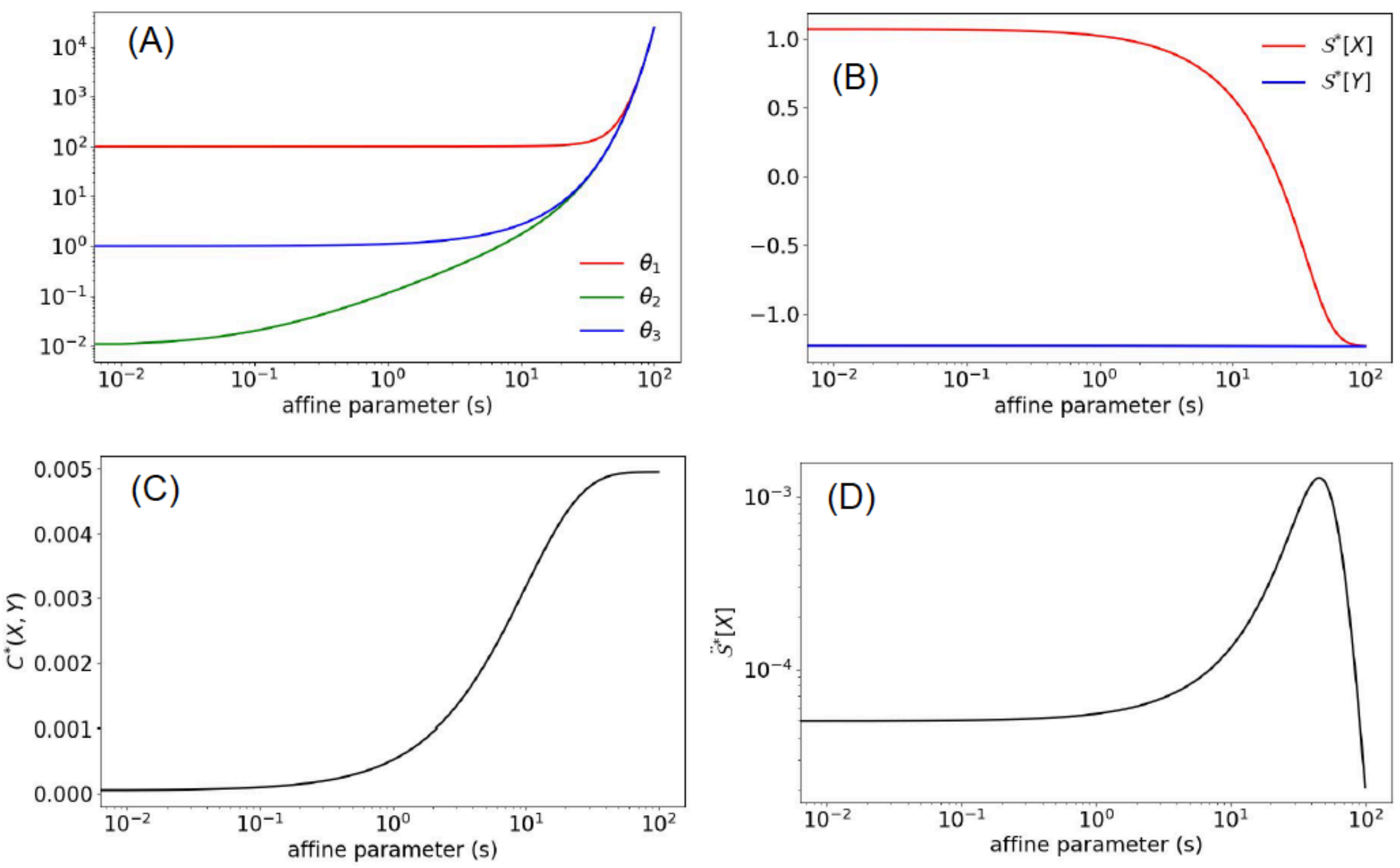

**Fig. 1** The numerical results under the initial conditions: $\theta^1(0)=100$, $\theta^2(0)=0.01$, $\theta^3(0)=1$ and $\dot{\theta}^1(0)=\dot{\theta}^2(0)=\dot{\theta}^3(0)=0.1$. (A) The log-log plots of the solution of the geodesic equation. (B) The semi-log plots of the subentropies $S^*[X]$ and $S^*[Y]$ with respect to *s*. (C) The semi-log plot of the covariance $C^*(X,Y)$ with respect to *s*. (D) The log-log plot of $\ddot{S}^*[X,Y]$ with respect to *s*. All quantities are dimensionless.

In Fig. 2, the initial correlation is negative (e.g. anticorrelation). Again, the result shows adaptation. The behaviors similar to those in Fig. 1 (B) and (D) are recognized. However, $C^*(X,Y)$ exhibits the antisigmoidal behavior, in contrast to Fig. 1 (C).

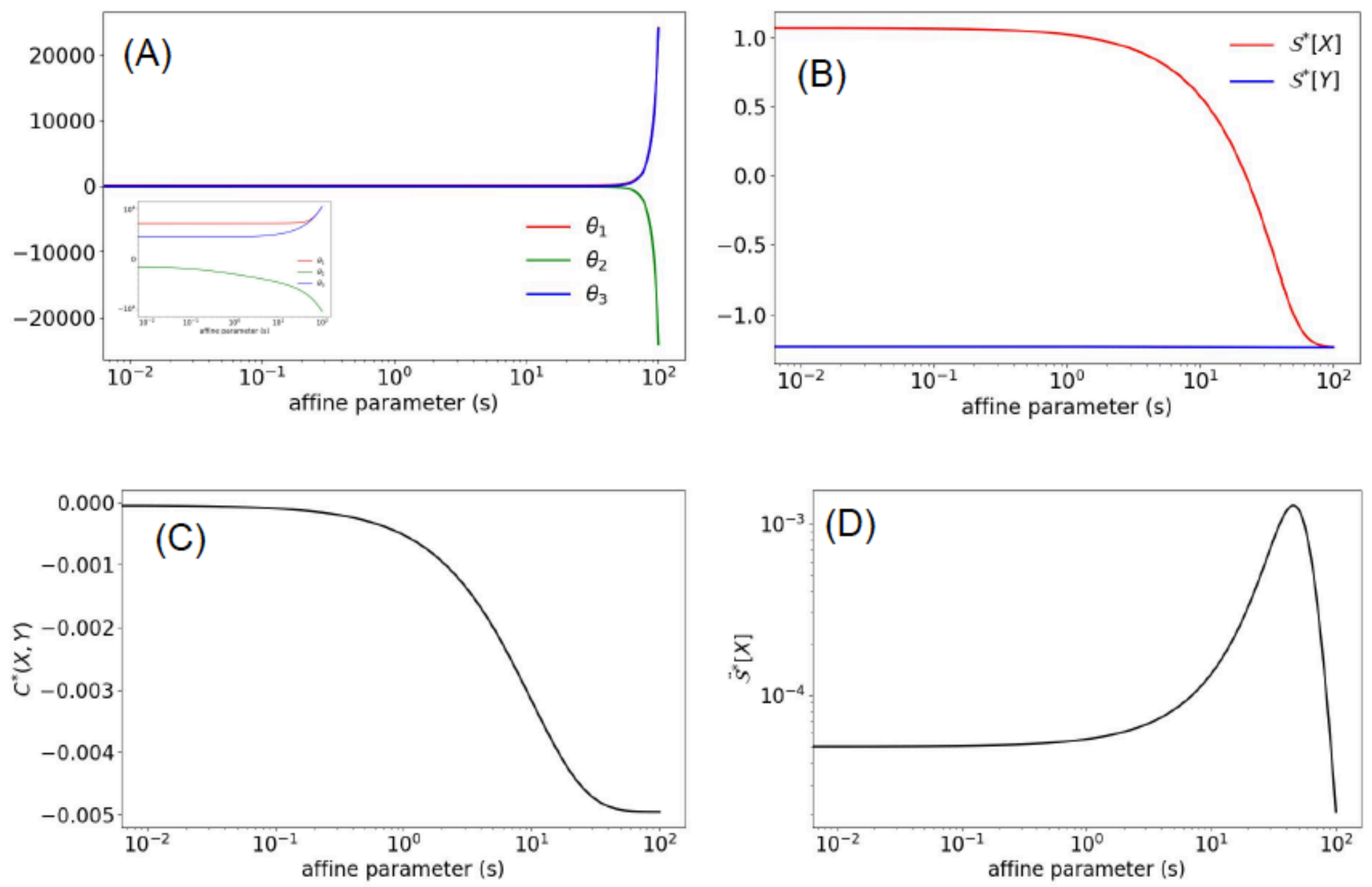

**Fig. 2** The numerical results under the initial conditions: $\theta^1(0)=100$, $\theta^2(0)=-0.01$, $\theta^3(0)=1$, $\dot{\theta}^1(0)=0.1$, $\dot{\theta}^2(0)=-0.1$ and $\dot{\theta}^3(0)=0.1$. (A) The log-log plots of the solution of the geodesic equation (inset: magnification). (B) The semi-log plots of the subentropies $S^*[X]$ and $S^*[Y]$ with respect to *s*. (C) The semi-log plot of the covariance $C^*(X,Y)$ with respect to *s*. (D) The log-log plot of $\ddot{S}^*[X,Y]$ with respect to *s*. All quantities are dimensionless.

**(3-2) Analytic solution in the degenerate case:** It turns out to be possible to analytically solve Eqs. (33)-(35) in the special case when

$$\ddot{S}^*[X]=0. \tag{39}$$

This implies that the Wronskian in Eq. (36) vanishes and therefore $\theta^1$ and $\theta^2$ are linearly dependent (i.e. degenerate). Actually, it does not lead to processes of interest, as will be seen. The degeneracy allows us to set $\theta^2 = c\theta^1$, where $c$ is a nonzero constant. Substituting this into Eqs. (33) and (34), we have $\ddot{\theta}^1 / \dot{\theta}^1 = \dot{\theta}^1 / \theta^1$ and $\ddot{\theta}^2 / \dot{\theta}^2 = \dot{\theta}^2 / \theta^2$, which are solved respectively as $\theta^1(s) = \theta_0^1 e^{\kappa s}$ and $\theta^2(s) = \theta_0^2 e^{\kappa s}$, where $\theta_0^i \equiv \theta^i(0)$ and $\kappa$ is a nonzero constant. Then, using these solutions in Eq. (35) and setting $\theta^3 = \varphi e^{\kappa s}$ with $\varphi(0) = \theta_0^3$, we have $\ddot{\varphi} / \dot{\varphi} = \theta_0^1 \dot{\varphi} / [\theta_0^1 \varphi - (\theta_0^2)^2]$. The solution of this equation is given by $\varphi(s) = \left\{ (\theta_0^2)^2 + [\theta_0^1 \theta_0^3 - (\theta_0^2)^2]\, e^{\lambda s} \right\} / \theta_0^1$, where $\lambda$ is a constant. Therefore, we have $\theta^3(s) = \left\{ (\theta_0^2)^2 e^{\kappa s} + [\theta_0^1 \theta_0^3 - (\theta_0^2)^2] e^{(\kappa+\lambda)s} \right\} / \theta_0^1$. Thus, the geodesic equation has analytically been solved under the condition in Eq. (39). The initial values, $\theta_0^i$'s, are chosen in such a way that the requirements in Eqs. (16) and (17) are fulfilled. Then, from the learning condition in Eq. (29), it follows that $\kappa + \lambda > 0$ ($\kappa, \lambda \neq 0$). Now, let us quantify the correlation between $X$ and $Y$ using the mutual information. Using the analytic solutions, we obtain $I^*[X;Y] = (1/2)\ln\left\{ 1 + \left( (\theta_0^2)^2 / [\theta_0^1 \theta_0^3 - (\theta_0^2)^2] \right) e^{-\lambda s} \right\}$. As $s$ grows, this tends to linearly diverge if $\lambda < 0$, to exponentially decrease if $\lambda > 0$ and remains constant if $\lambda = 0$ (the total degeneracy, i.e. all $\theta^i$'s being degenerate). This behavior is irrelevant to the processes of learning and adaptation.

Thus, we conclude that Eq. (37) should hold for geodesic learning in the present model.

## 4. Summary and remarks

In order to study learning processes, we have considered a composite system of a decision-maker and the environment surrounding it and have described its state by the probability distribution characterized by a set of parameters. We have introduced the concepts of the information bath and the specific information, which make it possible to discuss learning processes in analogy with thermodynamics. Then, we have formulated geodesic learning based on Riemannian geometry with the Fisher-Rao metric associated with the state change represented by the parameters. Geodesic learning gives the geometrically-optimal process, comparison with which may show how a given process is efficient. In this theory, no kinetic equations are assumed for evolution of the distribution function, and the system is not governed by *a priori* dynamics but follows the strategy of geometric optimality. We have presented the bivariate Gaussian model as an illustrative example and have demonstrated how learning and adaptation are simultaneously observed.

The terminology "geodesic learning" has appeared in the literature of machine learning and psychology with its roots in the utilization of geometric constraints. In the field of machine learning, it seeks to harness the inherent geometric features of data [9] that are either physical characteristics or abstract manifold structure of the data [10], in order to improve algorithmic efficiency and accuracy. In psychology, it introduces an approach to self-directed learning, implying that learning trajectories should align with cognitive pathways of an individual for more intuitive and effective knowledge acquisition [11]. The ideas outlined in the present work manifest the potential of geodesic learning for the design of self-supervised learning systems, enabling them to learn autonomously by identifying and following the natural contours in the information landscape, and may be useful for in the fields of machine learning and psychology.

Cells, as complex biological units, function as decision-makers that continuously interpret and respond to their environments. This dynamic interaction leads to alterations in their internal states, notably in gene expression patterns. Recent advancements have revealed that the geometry of genetic expression is hyperbolic in nature [12]. Hyperbolic geometry, which is the first example of non-Euclidean geometry discovered by Bolyai and Lobachevsky, is categorized in Riemannian geometry. Understanding of cellular behaviors and cell decision-making, in particular, may significantly be enhanced if geodesic learning presented here and its appropriate generalizations are used as the case of reference. It will also allow the coarse-grained description of gene regulation in analogy with thermodynamics.

**Acknowledgement**

AB would like to thank Tata Institute of Fundamental Research Hyderabad for the fellowship.

**Author contributions**

SA has framed the research and has written the manuscript. AB has done the numerical analysis. All of the authors have examined the theory and have equally contributed to improvement and completion of the manuscript.

**Declaration of competing Interest**

The authors declare that they have no known competing financial interests or personal relationships that could have appeared to influence the work reported in this paper.

## Data availability

This paper does not report either data generation or analysis.

## Funding

The work of HH has been supported by the "Life?" program (96732) of Volkswagenstiftung and the grant from the Bundes Ministeriums für Bildung und Forschung under agreement No. 031L0237C (MiEDGE project/ERACOSYSMED). He has also been supported by the RIG-2023-051 grant from Khalifa University and the AJF-NIH-25-KU grant from the NIH-UAE collaborative call 2023. SA has been supported in part by the grant from National Natural Science Foundation of China (No. 12375031), the Program of Fujian Province, China, and the Kazan Federal University Strategic Academic Leadership Program (PRIORITY-2030).

## References

[1] W.S. Bialek, Biophysics: Search for Principles, Princeton University Press, Princeton, 2012.

[2] E.L. MacLean, et al., How does cognition evolve? phylogenetic comparative psychology, Anim. Cogn. **15** (2012) 223-238, doi:10.1007/s10071-011-0448-8.

[3] M. van Dujin, Phylogenetic origins of biological cognition: convergent patterns in the early evolution of learning, Interface Focus **7** (2017) 20160158,

doi:10.1098/rsfs.2016.0158.

[4] S.J. Maybank, The Fisher-Rao metric, Mathematics Today (December, 2008) 255-257.

[5] A. Barua, J.M. Nava-Sedeño, M. Meyer-Hermann M, H. Hatzikirou, A least microenvironmental uncertainty principle (LEUP) as a generative model of collective cell migration mechanisms, Sci. Rep. **10** (2020) 22371, doi: 10.1038/s41598-020-79119-y.

[6] C. Ou, S. Abe, Exotic properties and optimal control of quantum heat engine, EPL 113 (2016) 40009, doi: 10.1209/0295-5075/113/40009.

[7] T.M. Cover, J.A. Thomas, Elements of Information Theory, Wiley, NY, 2006.

[8] P. Petersen, Riemannian Geometry, third ed., Springer International, Cham, 2016.

[9] A. Attyé A, et al., TractLearn: a geodesic learning framework for quantitative analysis of brain bundles, NeuroImage **233** (2021) 117927, doi:10.1016/j.neuroimage.2021.117927.

[10] P.T. Fletcher, Geodesic regression and the theory of least squares on Riemannian Manifolds, Int. J. Comput. Vis. **105** (2013) 171-185, doi:10.1007/s11263-012-0591-y.

[11] C.M. Leaf, B. Louw, I. Uys, The development of a model for geodesic learning: the geodesic information processing model, Die Suid-Afrikaanse Tydskrif vir Kommunikasieafwykings, **44** (1997) 53–70.

[12] Y. Zhou, T.O. Sharpee, Hyperbolic geometry of gene expression, iScience **24** (2021) 102225, doi: 10.1016/j.isci.2021.102225.